# Triple Coding Empowered FDMA-CDMA Mode High Security CAOS Camera

Nabeel A. Riza, [1,*] and Mohsin A. Mazhar[1]

[1] *Photonic Information Processing Systems Laboratory, School of Engineering, University College Cork, College Road, Cork, Ireland*
*Corresponding author: n.riza@ucc.ie*

For the first time, the hybrid triple coding empowered Frequency Division Multiple Access (FDMA)-Code Division Multiple Access (CDMA) mode of the CAOS (i.e., Coded Access Optical Sensor) camera is demonstrated. Compared to the independent FDMA and CDMA modes, the FDMA-CDMA mode has a novel high security space-time-frequency triple signal encoding design for robust, faster, linear irradiance extraction at a moderately High Dynamic Range (HDR). Specifically, this hybrid mode simultaneously combines the linear HDR strength of the FDMA-mode Fast Fourier Transform (FFT) Digital Signal Processing (DSP)-based spectrum analysis with the high Signal-to-Noise Ratio (SNR) provided by the many simultaneous CAOS pixels photo-detection of the CDMA-mode. In particular, the demonstrated FDMA-CDMA mode with P FDMA channels provides a P times faster camera operation versus the equivalent linear HDR Frequency Modulation (FM)-CDMA mode. Visible band imaging experiments using a Digital Micromirror Device (DMD)-based CAOS camera operating in its passive light mode demonstrates a P=4 channels FDMA-CDMA mode, illustrating high quality image recovery of a calibrated 64 dB 6-patches HDR target versus the CDMA and FM-CDMA CAOS modes that limit dynamic range and speed, respectively. For the first time, demonstrated is the simultaneous dual image capture capability of the FDMA-CDMA mode using Silicon (Si) and Germanium (Ge) large area point photo-detectors allowing the capture of the Ultraviolet (UV) – Near Infrared (NIR) 350-1800 nm full spectrum. The active FDMA-CDMA mode CAOS camera operation is also demonstrated using P=3 LED light sources, each with its unique optical spectral content driven by its independent FDMA frequency. This illuminated target spectral signature matched active CAOS mode allows simultaneous capture of P images without the use of P time multiplexed slots operation tunable optical filter. Applications for such a FDMA-CDMA camera includes controlled light illumination food inspection to bright light exposure security systems.

## 1. INTRODUCTION

There are applications in industry and science across the UV, visible and NIR wavelength range that can benefit from a high brightness capability, high security, linear irradiance extraction, moderately HDR (e.g., > 60 dB), robust SNR, full spectrum (e.g., 350 nm to 1800 nm) operations camera unit [1-6]. Most deployed camera systems use multiple image sensors, e.g., silicon CCD, silicon CMOS and InGaAs IR Focal Plane Array (FPA) direct pixel readout sensors along with multiple wavelength filters and optics to realize a multi-spectrum imaging system [7]. Although such commercially available cameras have good weak light sensitivity and have been engaged across various applications, they still face certain limitations such as high costs, cooling requirements, and non-linearities that restrict high SNR and linear HDR pixel irradiance extraction, particularly in the non-silicon detection bands [8-11]. In addition, image pixel photo-charge is collected in a non-secure way as DC voltage levels from the multi-pixel array in the sensor chip with this directly read out output being susceptible to eavesdropping and manipulations.

As far back as 1949, a spinning on/off shutter-type disks-based optical coding of light with point detector-based light capture was realized for infrared spectrometry [12]. Subsequent follow-on works in the late 1960's and beyond showed that indeed simultaneous photo-detection using on/off coding of Q optical spectra channels versus a single spectra channel provides a $\sqrt{(Q/2)}$ advantage in detection SNR [13-14]. With the availability of the DMD for on/off spatial coding of light in the late 1990s, several optical architectures emerged for both spectrometry [15-16] and imaging [17] including in 2001 for point detector-based agile pixel imaging [18-19] and in 2006 for the implementation of compressive imaging algorithms [20-21] that is popularly being called *Single Pixel Imaging*. Unlike these prior works, a linear HDR highly programmable full spectrum optical imaging instrument recently proposed is the CAOS smart camera [22-23] that can lead to next generation hardware and software flexible high security "thinking" cameras [24] for application adaptive operations with extreme up-to 177 dB linear dynamic ranges such as demonstrated in ref.25 [25]. The full spectrum coverage of 320 nm to 2700 nm range via the CAOS camera is possible because of the broadband operation of the TI DMD that is deployed in the camera. As described in detail in the earlier CAOS camera works [22-25], the linear HDR capability of the CAOS camera is possible because of the time-frequency Radio Frequency (RF) wireless multi-access phone network style image pixel irradiance encoding and DSP-based time-frequency linear HDR low noise spectrum analysis and correlation decoding and inter-pixel crosstalk filtering. Specifically, high Mega-samples per second Analog-to-Digital Converter (ADC)-based digitization of the CAOS encoded photo-detected signal combined with DSP-based high spectral gain filtering enables linear HDR normalized pixel irradiance recovery, much like RF receivers in phone and other mobile units. In addition, intelligent spatial instantaneous sampling of the image plane pixels using RF spectral encoding provides a mechanism for inter-pixel as well as inter-image (i.e., when using active light sources) crosstalk control via RF spectral filtering. One attractive mode of the CAOS camera described and demonstrated in detail in an earlier work is the hybrid FM-CDMA mode [26] that has the capability to provide pixel irradiance extraction with both linear HDR and high SNR. The CDMA-mode simultaneously extracts many pixels (e.g., Q=3600 pixels) providing a higher light level for detector-noise limited photo-detection giving the classic $\sqrt{(Q/2)}$ SNR advantage over one pixel at a time photo-detection, as analyzed and proven in prior-art experiments in the late 1960s and thereafter [13-14, 27-30]. In addition, the FM-mode within the CDMA-mode implies that each CDMA signaling bit in the time sequence signal encoding for a given pixel is provided with an RF carrier so one can engage low noise DSP FFT spectral processing for linear HDR irradiance pixel recovery. Another useful CAOS mode when using FM-style coding is the FDMA-Time Division Multiple Access (TDMA) mode as multiple frequencies within each TDMA time slot that is used for scanning the image pixel zone can also be used for creating a faster access to complete the imaging operation. Details of the digital FDMA-TDMA mode has been described and demonstrated in detail in ref.31 [31].

The purpose of the present paper is to present for the first time, the design and experimental demonstration of the combined hybrid design novel FDMA-CDMA mode. When compared to the previously demonstrated CDMA, FM-CDMA, and FDMA-TDMA modes, the novel hybrid FDMA-CDMA mode is more powerful as it simultaneously empowers pixel irradiance extraction with four unique features, namely, high security, linear HDR, high SNR, and high speed, particularly in the context of simultaneously imaging from the UV to the near IR band. High SNR comes via the CDMA mode, linear HDR comes via the FM mode using DSP gain, while the CDMA and FDMA modes combined together reduces the CDMA time code length in the FDMA-CDMA mode to allow faster image acquisition. High security of the captured image data comes from the natural triple protection of the space-time-frequency signaling embedded in the FDMA-CDMA mode. Experiments for the visible band using a 64 dB 6-patch calibrated target successfully demonstrates core capabilities using 4 FDMA channels for the FDMA-CDMA mode and shows correct image recovery vs the FM-CDMA mode. In addition, for the first time demonstrated is the FDMA-CDMA mode that simultaneously provides two independent images, one covering the silicon band from 350-1000 nm and a second image covering 800-1800 nm germanium detection band, delivering full spectrum imaging. In addition, using 3 independent spectrum LED sources for FDMA encoding, the active FDMA-CDMA mode is demonstrated that simultaneously provides 3 target images for the 3 different optical spectra. The FDMA-CDMA mode CAOS camera operation can create impact across applications such as food inspection and remote sensing requiring full spectrum linear HDR with adequate SNR near UV-NIR extractions for robust materials measurements and secure vision. The rest of the paper describes the details of the CAOS camera FDMA-CDMA design and experimental demonstrations.

## 2. FDMA-CDMA MODE CAOS CAMERA DESIGN

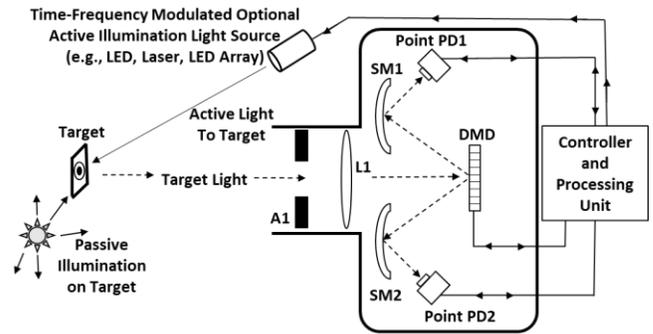

Fig. 1. Top view of the CAOS camera design for FDMA-CDMA mode operations for dual simultaneous spectral band operations. SM1/SM2: Spherical Mirrors; A1: Aperture; PD1/PD2: Different Material Large Area Point Detectors.

Fig.1 shows the top view of the CAOS camera optical design for FDMA-CDMA mode dual simultaneous spectral band operations. Note that the active and passive illumination mode options of the CAOS camera that was proposed earlier in ref.23 is also deployed in the Fig.1 design [23]. Depending on the optical spectral properties of the passive and customized active controlled illumination sources as well as the illuminated target spectral signatures, optimized input spectrum light from a target scene enters the controlled aperture A1 to pass through an imaging lens system L1 to form an image on the DMD time-frequency CAOS encoding plane. The size of the open aperture A1 can be used to optimize reduced aberration effects as well as limit the camera Field of View (FOV) and light levels incident on the DMD. L1 has an F1 focal length and forms a demagnification system between the scene plane and the DMD plane. The +1 (i.e., +θ) tilt micro-mirror state of the DMD sends light from the DMD plane to the spherical mirror SM1 of focal length SF1 to be imaged on to a large area high speed point detector PD1. The -1 (i.e., -θ) tilt micro-mirror state of the DMD sends light from the DMD plane to the spherical mirror SM2 of focal length SF2 to be imaged on to a large area high speed point detector PD2. One feature of the Fig.1 optical design is that SF1 and SF2 focal lengths each can be optimized to design appropriate demagnifications between the DMD region of interest for imaging and the specific and possibility different active areas of the large area PD1 and PD2 detectors. For instantaneous full spectrum imaging, ideally PD1 and PD2 photo-responses cover independent non-overlapping spectral bands tailored by using the optimal photo-sensitive materials in combination with associated fixed or programmable optical filters that can be placed before the PD1 and PD2, respectively. The two photo-detected CAOS encoded signals coming from PD1 and PD2, each carrying its specific spectrum image data are amplified and sampled by a pair of high speed ADCs. These two digitized signals are next sent for time-frequency (Hz domain) DSP via the camera control and processing electronics that generate two instantaneous spectral images of the scene independently provided by PD1 and PD2. For example, if PD1 is UV band sensitive and PD2 is NIR band sensitive, independent UV and NIR images of the observed target can be captured at the same time using the FDMA-CDMA CAOS mode providing robust SNR with adequate HDR linear irradiance capture at faster speeds versus the similar FM-CDMA mode.

Note that the CDMA mode uses a special coding method with an Error Correction (EC) bit added to the Walsh code to enable simultaneous capture of both PD1 and PD2 provided images [32]. On the contrary, the FM-CDMA and FDMA-CDMA modes do not require the EC bit overhead as its FM carrier phase independent FFT DSP spectrum analysis can directly encode the DMD viewed image map and can also then directly decode the 2 independent images produced from PD1 and PD2. This special property is possible in the CAOS camera due to the fact that the digital on/off two tilt-states micromirror setting natural to FM encoding of the imaged pixel irradiance produces the optical FM signals at both PD1 and PD2, allowing for CAOS encoded simultaneous image extractions via DSP from both PD1 and PD2. This CAOS camera feature also inherent in the presented FDMA-CDMA mode is highly useful for linear HDR dual image extractions over different spectral bands when using two PDs with different spectral responses.

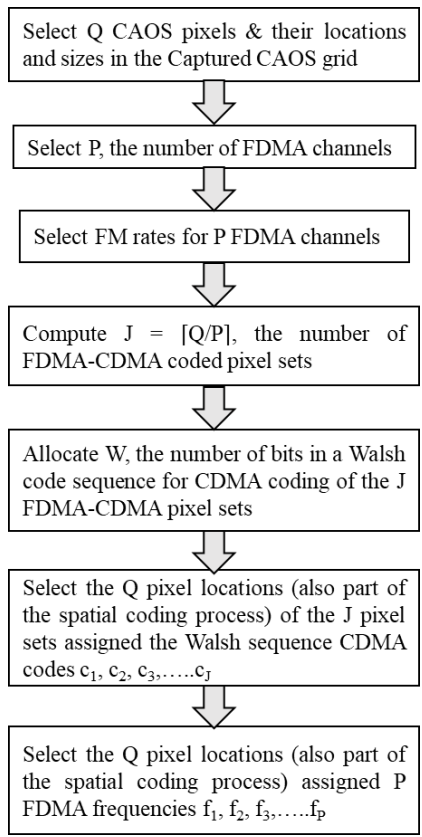

Fig.2. Flow chart for the CAOS space-time-frequency signaling FDMA-CDMA encoding operation per image frame.

Fig.2, Fig.3 and Fig.4 are collectively provided to illustrate the implementation of the FDMA-CDMA mode within the CAOS camera. Fig.2 and Fig.4 show the flow chart for the CAOS signal processing encoding and decoding operations, respectively. Fig.2 flow chart highlights the order of operations for the FDMA-CDMA CAOS encoding process. Note that any non-regular partially filled sparse image grid space can also be used for CAOS imaging which allows efficient selected pixels specific irradiance extraction. The CAOS pixels within any given FDMA pixel set can be chosen from any pixel location in the image space providing a space coding mechanism in addition to the already present time-frequency CAOS encoding provided by the FDMA-mode combined with the CDMA-mode. Thus, a natural feature of the presented FDMA-CDMA CAOS mode is the inherent security of the harvested encoded image as only the user knows which CAOS pixels in the image frame have been selected to receive a specific FM frequency as well as a specific CDMA code, thus protecting the image pixel irradiances by a triple coding method combining space-time-frequency aspects of signaling within the encoder. The specific FDMA-CDMA encoding steps per CAOS image frame that are also highlighted in Fig.2 are as follows:

1. Select the number of Q CAOS pixels as well as their locations and sizes in the Captured CAOS grid on the DMD plane.
2. Select P, the number of FDMA channels to be used within the Q CAOS pixels.
3. Select the different FM rates to be deployed for the P FDMA channel set.
4. Compute J = ⌈Q/P⌉, the number of FDMA-CDMA coded pixel sets. The symbol with half brackets ⌈ ⌉ represents the ceiling function.
5. Given J, allocate W, the number of bits in a Walsh code sequence deployed for CDMA coding of the J FDMA-CDMA pixel sets.
6. Select the Q pixel locations (part of the spatial coding process) of the J pixel sets assigned the different W-bits Walsh sequence CDMA codes $c_1, c_2, c_3, \ldots c_J$.
7. Select the Q pixel locations (also part of the spatial coding process) assigned different FDMA frequencies from the P set of frequencies $f_1, f_2, f_3, \ldots f_P$.

Following these steps 1 to 7, the space-FDMA-CDMA encoding process is completed so that each unique CAOS pixel location in the Q-pixels grid also has a unique CDMA code as well as a unique FM rate from the FDMA frequencies set. Fig.3 shows the CAOS camera FDMA-CDMA mode image signal encoding example operational design. Specifically, Fig.3(a) shows the allocation of P FDMA frequency channels and J CDMA codes using a basic raster-line space coding allocation approach across the Q= M × N CAOS pixels rectangular grid on the DMD where n=1,2, 3,.., N and m=1,2,3,..,M along the image grid vertical and horizontal axes, respectively. Here the mn$^{th}$ CAOS pixel irradiance is $I_{mn}$ and is given a specific CDMA, FDMA coding pair set, e.g., $c_2, f_2$ for the m=1, n=2 CAOS pixel with an irradiance of $I_{12}$ per Fig.3(a) coding matrix. In general, both FDMA and CDMA codes are functions of time, i.e., c(t) and f(t). The typical c(t) Walsh encoding sequence is a 1 or 0 digital value W bits sequence (see Fig.3(b)) with the sequence individual bit time of T seconds, i.e., t=wT with w=1,2, 3,…W. The simple raster line scan spatial coding approach in Fig.3(a) is shown as an illustrative example as it is also used in the first experimental demonstration later in the paper. Note that advanced algorithms including machine learning based methods can be deployed for placement of the P FDMA and J CDMA codes across the Q CAOS pixels grid. This space-time-frequency coding allocation over the observed image grid can also be updated over pre-determined time positions and scales, including over varying number of consecutive imaging frames giving an additional level of complexity and security to the CAOS camera.

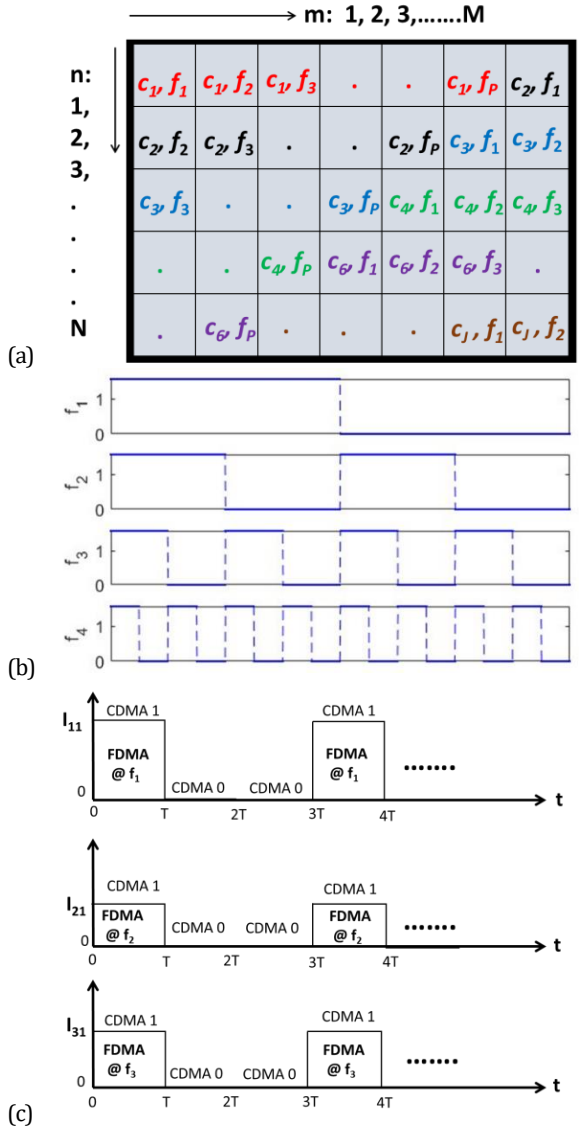

Fig. 3 CAOS camera FDMA-CDMA mode image signal encoding operational design example. (a) Allocation of P FDMA frequency channels and J CDMA codes using a raster-line space coding allocation approach across the Q = M × N CAOS pixels grid on the DMD. (b) Design of P=4 FDMA channel frequencies $f_1$, $f_2$, $f_3$ and $f_4$ using square wave signals time-frequency modulation. (c) Shown is the partial 1,0,0,1 CDMA code time bits sequence for the first 3 CAOS pixels in the top line of the image in Fig.3(a) with irradiances $I_{11}$, $I_{21}$, and $I_{31}$ coded with FDMA channel frequencies $f_1$, $f_2$, and $f_3$, respectively.

Fig.3(a) also represents the example triple coding matrix deployed by the CAOS camera. This coding matrix has a matrix element $d_{mn}$ at the $mn^{th}$ image pixel location given as a product of the assigned W bits Walsh CDMA code sequence $c_{mn}$ and the assigned FDMA frequency signal $f_{mn}$ that can be the same from image frame to image frame across all J assigned CDMA codes. Note that assigned P FM frequencies $f_1$, $f_2$, $f_3$,...,$f_P$ can also be hopped amongst the P frequencies set from CDMA bit time to next bit time within a W time bits frame encoding process, adding an additional level of security to the encoded signal produced by the CAOS camera. In other words, the assigned coding matrix FM signal allocations changes in the CDMA W bit sequence and up-to W different coding matrices are required for proper CAOS imaged frame encoding and decoding indicating a higher level of image security. Apart from within imaged frame FDMA frequency allocation changes, a further additional level of security can be added by implementing changes in the allocated J CDMA codes from frame to frame. Such camera security capabilities are possible due the intrinsic nature of the FDMA-CDMA mode of the CAOS camera.

Next, as an illustration, Fig.3(b) shows a sample time window of the design of P=4 FDMA channel frequencies $f_1$, $f_2$, $f_3$ and $f_4$ using square wave FM signals inherent to DMD time-frequency modulation when using the passive illumination mode. The fundamental carrier is $f_1$ Hz and the allocated FDMA $p^{th}$ channel frequency $f_p=2^{p-1} f_1$ with p=1,2,..P so that inter-channel CAOS pixels crosstalk is kept to a minimum due to RF odd harmonics spectral crosstalk given the square wave nature of assigned DMD implemented FDMA carriers. In addition, to ensure complete carrier cycles within a CAOS pixel FDMA encoding time window T, $f_1$=k $\Delta f$ where k=1, 2, 3, ... with $\Delta f$=1/T. Note that when using the active illumination mode, the external light sources such as LEDs and lasers can also provide sine wave FM light modulations giving a higher degree of FDMA carrier selection flexibility leading to minimal harmonics and closer placement of FDMA frequencies. In addition, use of higher (e.g., 30 KHz) FM rates is possible versus available from current DMDs limited to 11 KHz. Also note that higher FM rates leads to lower 1/f electronic noise in the signal processing electronics where f is frequency.

Fig.3(c) shows an example first four CDMA code bits sequence, namely, 1,0,0,1 at a CDMA bit rate of $f_B$=1/T bps for the first 3 CAOS pixels in the top line of the Fig.3(a) image with irradiances $I_{11}$, $I_{21}$, and $I_{31}$ coded per Fig.3(a) coding matrix with FDMA channel frequencies $f_1$, $f_2$, and $f_3$, respectively. Recall that the P FDMA frequencies can be allocated to any choice of P CAOS pixels that all use an assigned CDMA encoding bit sequence versus a line-based allocation as shown in Fig. 3(a) where P CAOS pixels follow a raster line coding allocation. At any CDMA bit coding window of T duration for a specific CDMA code, at most P CAOS pixels covering P FDMA frequencies simultaneously encode the imaged scene versus a single CAOS pixel when using the FM-CDMA CAOS mode. In effect, the P channel FDMA-CDMA mode can be designed to provide a much faster camera encoding operation versus FM-CDMA while still ensuring all Q CAOS pixels are simultaneously detected at the PDs for high $\sqrt{(Q/2)}$ SNR operations similar to the CDMA-mode for Q CAOS pixels. Recall from Fig. 2 encoding flow chart that if Q/P is not a whole integer, one picks the nearest whole integer J higher than the Q/P fractional number to design the required number of J FDMA-CDMA coded CAOS pixel sets. This computation employs the ceiling function so J= ⌈Q/P⌉. For example, with M=55, N=37, Q=M×N=55×37=2035, P=8, Q/P=2035/8=254.375, one gets J=⌈Q/P⌉=255. In effect, there will be 254 sets of P=8 simultaneous CAOS pixels using all P=8 FDMA frequencies per set covering 254 × 8=2032 CAOS pixels. Given there are a total Q= 2035 CAOS pixels, the remaining 3 CAOS pixels creates a 255th FDMA-CDMA encoded pixels set. This final set has 3-pixels and uses for example only the first 3 of the 8 available FDMA frequencies for CAOS encoding. To achieve CDMA coding for J pixel sets, a W bits orthogonal Walsh codes sequence is found and deployed where W ≥ J. In the case J=255, W=256 bits Walsh sequence can be used. As a comparison, FM-CDMA CAOS imaging for the Q=2032 CAOS pixels would require W=2048 bits Walsh code time sequences versus the W=256 bits for a P=8-channel FDMA-CDMA mode indicating a 2048/256=8 times faster CAOS frame encoding time highlighting the speed advantage using the P=8 FDMA channels count.

Mathematically, one can express the photo-detected current at the $w^{th}$ bit sequence of the CDMA code within the FDMA-CDMA mode encoded CAOS camera as:

$$i(w,t) = G \sum_{n=1}^{N} \sum_{m=1}^{M} I_{mn} d_{mn}(w,t) \quad (1)$$

Here w increases from 1, 2, 3,...., W, giving W photo-detected current readings provided by the PDs in the CAOS camera. G is a scaling gain factor incorporating various parameters such as photo-detector responsivity and electronic amplifier gain. In addition, the coding matrix element at the $mn^{th}$ image pixel location with optical irradiance $I_{mn}$ is given by:

$$d_{mn}(w,t) = c_{mn}(t) f_{mn}(w,t) \quad (2)$$

As mentioned earlier for higher security operations, the coding matrix elements $d_{mn}$ can change its FDMA frequency allocation $f_{mn}$ value from bit to bit within the encoded image frame. In addition, $d_{mn}$ can change its CDMA code allocation $c_{mn}$ from image frame to frame. In case the coding matrix does not change within the frame and also from frame to frame, $d_{mn}$ becomes independent of w and t with $d_{mn}(w,t) = d_{mn}$ and follows a spatial coding-only approach where specific pixel locations use a specific CDMA code and specific FDMA FM channel encoding during continuous camera operations.

Next, Fig. 4 shows the flow chart for the CAOS signal processing FDMA-CDMA decoding operations using the photo-detected signals in the CAOS camera. Specifically, for recovery of the 2 different spectrum Q pixel images observed by the CAOS camera using the proposed FDMA-CDMA mode and the different response PD1 and PD2, the two digitized ADC signals from PD1 and PD2 first undergo RF spectrum analysis on a CDMA bit by bit basis using the Eq. 1 signal with w=1,2,....., W. The RF spectrum analysis is done for example by DSP-based FFT. The ADC $f_S$ sampling rate and CDMA code bit rate $f_B$ are controlled to adjust the F-point FFT DSP gain of 10log(F/2) dB for spectrum analysis where F is number of data samples in CDMA bit duration T. The RF spectrum produced per time bit interval T contains P RF spectral peaks corresponding to the P FM encoding channels in the FDMA set. Given there are W bits in the CDMA sequence, there are W sets with each set containing P spectral peaks. These RF spectral readings in the W spectral sets contain the CDMA encoded irradiance information of the Q CAOS pixels. Next, all the FFT peak readings per bit time for a selected FDMA channel (e.g., f=$f_1$) are used to create a bit sequenced FFT peak data signal. This FDMA-decoded signal sequence data set at a specific FM channel next undergoes time-integrated correlation operations with the deployed J CDMA code sequences to produce scaled irradiance values for J CAOS pixels. This CDMA-decoding based correlation process is repeated with another FM channel FDMA-decoded signal sequence data set to produce a new set of J CAOS pixel scaled irradiance values. This process is continued for all W spectral sets till all Q scaled irradiances for the imaged Q pixels are recovered. As the CAOS camera authorized operator knows the scaled irradiance recovery assigned FDMA, CDMA spatial coding grid, the computed scaled irradiances can be allocated to the correct CAOS pixel locations in the harvested grid to produce the full recovered image with Q CAOS pixels. Given that the presented decoding process is implemented independently for both PD1 and PD2 photo-detected signals, two simultaneous Q-pixel images are recovered by the CAOS camera. Furthermore, each Q-pixel optimized optical spectrum centric image can be normalized using the brightest scaled irradiance in each image.

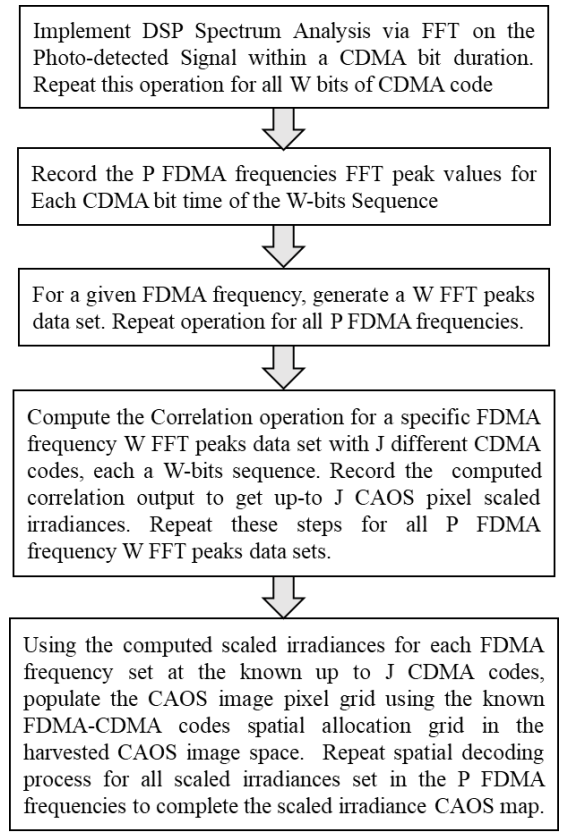

Fig. 4. Flow chart for the CAOS signal processing FDMA-CDMA decoding operation.

A feature of the active FDMA-CDMA mode is the operation of two different types of spatial target illumination modes based on the desired application for the CAOS camera. In one approach, all the different spectra FDMA encoded light beams overlap each other on the target. In this special case of beams overlap, each of the P FDMA encoded images has Q pixels and J is not equal to [Q/P]. Such an approach allows faster spectrally selective imaging of the target by simultaneously producing P different images for P different optical spectra, all without the need of a programmable spectral filter operating in a time multiplexed mode with P time slots. Note that the Fig.2 flow chart for this type of active FDMA-CDMA mode operation does not engage the 4$^{th}$ step in the encoding process as in this case, W is allocated for the larger count Q pixels as all P FDMA encoded images each have Q pixels. In another active illumination approach that is similar to the passive FDMA-CDMA mode, the different FDMA encoded beams illuminate different regions of the target that need pixel irradiance extraction, thus creating a Space-Division-Multiple-Access (SDMA) operation. In this case, all P optical spectra can also be the same for the P independent beams, creating a SDMA-FDMA-CDMA CAOS active mode that works faster than the first described active FDMA-CDMA beams overlap mode for Q pixels/image, but using the same basic Fig. 2 encoding flow chart as the passive FDMA-CDMA CAOS mode. This is the case because each SDMA beam can have up-to J CAOS pixels, so the encoding process follows the full Fig.2 flowchart. Also note that the SDMA-FDMA-CDMA active mode provides use of a higher RF spectrum bandwidth and positional flexibility of the sine-wave FM rates than possible with the DMD generated square wave signals-based passive FDMA-CDMA mode.

## 3. FDMA-CDMA MODE CAOS IMAGING EXPERIMENTS

The Fig.1 CAOS camera is assembled in the laboratory using the following components. Image Engineering LG3 40 Klux 400 – 800 nm white LED lightbox; Avantes AvaLIGHT-HAL-S-Mini Pro-lite 2850 K bulb color temperature, 4.5 mW power, 350 – 2500 nm spectrum, 600 μm diameter fiber feed with a 5 cm focal length 2.5 cm diameter collimation lens and a $2.1^0$ beam divergence; Vialux DMD model V-7001 with micromirror size of 13.68 μm × 13.68 μm; DELL 5480 Latitude laptop for control and DSP, National Instruments 16-bit ADC model 6211; Thorlabs components include multi-alkali 300 – 800 nm point PMT Model PMM02 with 20 KHz bandwidth as PD1 for first passive FDMA-CDMA mode experiment, Si 320 – 1100 nm point PD model PDA100A2 with electronic Variable Gain Amplifier (VGA) set to 30 dB for PD1 and Ge 800 – 1800 nm point PD model PDA50B-EC with 30 dB electronic VGA for PD2 for second passive FDMA-CDMA mode experiment, ~13 mm diameter Iris A1, 5.08 cm diameter uncoated broadband lenses L1 with F1=10 cm and a SM1/SM2 with SF1/SF2=3.81 cm. Key inter-component distances are: 11.7 cm between L1 and DMD; 64 cm between Avantes lens output plane and L1, 129 cm between LG3 output plane and L1, 10 cm between DMD and SM1/SM2 and 6.2 cm between SM1/SM2 and point PD1/point PD2. Fig.5 shows a 2 × 3 patch HDR calibrated transmissive target using the LG3 lightbox illumination for the first CAOS camera experiment with the novel passive FDMA-CDMA mode. The 64, 58, 48, 30, 20 DR dB values of the 6 patches target with a 1.08 cm diameter patch size are optimized using different Thorlabs ND filters. The 0 dB patch is an open patch with no ND filters.

To image the Fig.5 target, the DMD is programmed to generate a Q=44 × 29 =1276 CAOS pixels grid where each CAOS pixel is 8 × 8 micromirrors in size. The point PMT is used as PD1 to capture the white light from LG3. To implement the passive FM-CDMA mode, a FM frequency of 1024 Hz is deployed along with a 1280 bits Walsh code for CDMA encoding with an $f_B$=1 Hz giving a total encoding time of 1280 sec. The ADC sampling rate is 65536 sps giving an FFT DSP gain of 45.16 dB with F=65536. Fig. 6 (left image) shows the captured target image in log scale for DR values in dB indicating that the FM-CDMA mode has successfully imaged 4 patches up to a DR=48 dB. Specifically, by taking a spatial average over the patch regions, the measured DR values are determined to be 0 dB, 20.5 dB, 29.3 dB and 48.6 dB with the weakest 48 dB patch measured at an SNR=3.3. The remaining 58 and 64 dB weaker light patches fail recovery with SNR < 1.

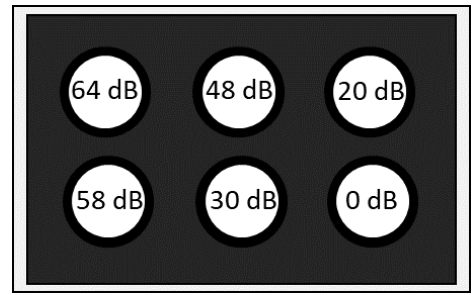

Fig.5 White light LED lightbox designed 6-patch 64 dB HDR test target.

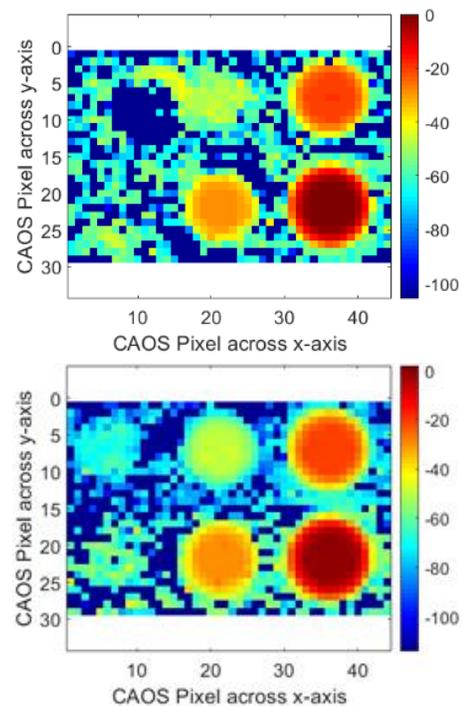

Fig. 6. (Top image) Passive FM-CDMA mode and (Bottom image) passive FDMA-CDMA mode CAOS camera measured DR dB scale images of the Fig.5 64 dB HDR target.

To demonstrate the higher linear HDR recovery of an image with improved SNR conditions and a 4 times faster encoding time for patch measurements, the proposed passive FDMA-CDMA mode is deployed using P=4 FDMA channels with $f_1$=128 Hz, $f_2$=256 Hz, $f_3$=512 Hz, $f_4$=1024 Hz. There are Q/P=1276/4=319=J sets of 4 simultaneous CAOS pixels using all 4 FDMA frequencies per set covering 319 × 4=1276 CAOS pixels. Note that a shorter W=320 bits Walsh code is used for CDMA encoding giving a CAOS image 4 times shorter encoding time of 320 s. In addition, 4 times fewer CAOS pixels are encoded by the CDMA codes, reducing the burden on the correlation-based decoding to decipher between the harvested irradiance values of the CAOS pixels and sharing part of the decoding burden with the FDMA encoding and spectrum analysis decoding operations. In effect, improved SNR is decoding follows with improved interpixel isolation and lower inter-pixel crosstalk during decoding signal processing. Fig. 6 right image shows that the FDMA-CDMA mode when compared to the FM-CDMA mode indeed has these capabilities as it has accurately recovered all 6 patches of the target with measured readings of 0 dB, 20.4 dB, 29.2 dB, 49.6 dB, 59.1 dB, and 64.1 dB with SNR ≥ 1.

As mentioned in Section 2, compared to the CDMA-mode, the FM-CDMA and FDMA-CDMA modes do not require the additional EC bit encoding and decoding step as the FM carrier phase independent FFT DSP spectral peak detection decoding step for both the FM-CDMA and FDMA-CDMA modes can directly decode the photo-detected signals from the two different PDs to deliver the 2 independent images. To demonstrate this special power of the FDMA-CDMA mode for full spectrum UV-NIR imaging, the Si and Ge point detectors are used for PD1 and PD2, respectively. For this second experiment, the UV-NIR Avantes 600 μm fiber is imaged using a M × N = 65 × 63 CAOS pixels grid with pixel size of 1 × 1 micromirrors, $f_B$=4 Hz, Walsh code W=1280 bits, encoding time of 420 sec, FFT DSP gain of 39.13 dB and FDMA and ADC settings per first experiment. Fig. 7 left image shows the 320-1000 nm band CAOS image with a measured DR=27.8 dB while the right image is the 800 – 1800 nm band CAOS image with a measured DR=31.05 dB. Both images are shown in linear scale and demonstrate successful simultaneous imaging of two spectral bands of the observed full spectrum fiber source target.

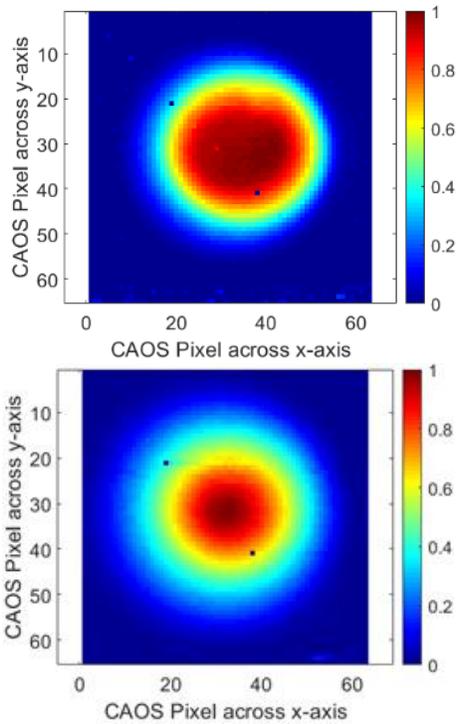

Fig.7. Top Image: 320-1000 nm CAOS image. Bottom Image: 800 – 1800 nm CAOS image. Both images shown in linear scale using passive FDMA-CDMA mode of the CAOS camera.

Next the CAOS camera assembled in the laboratory is modified to implement the overlapped beams active FDMA-CDMA mode using three Thorlabs model M series (M455L4-C1, M530L4-C1, M625L4-C1) LED collimated spot beam sources with different optical spectra. Specifically, beam 1, beam 2 and beam 3 have central wavelengths, 3-dB bandwidths parameters of (455 nm, 18 nm), (530 nm, 35 nm) and (625 nm, 17 nm), respectively. For a P=3 design active FDMA-CDMA implementation illuminating a two hole target, each LED is driven by a 5 V peak-to-peak sine wave signal on a 2.5 V DC bias level. The Green (G) LED1, Red (R) LED2 and Blue (B) LED3 are driven by FM sine waves with rates of $f_1$=25 KHz, $f_2$= 29 KHz, and $f_3$=35 KHz, respectively. To demonstrate the simultaneous capture of 3 spectrally selective images of the target when subjected to the 3 different LED spectral illuminations, combinations of 2 color filter sets are placed on the target holes. These optical filters are selected from a Thorlabs 3 color filter set with central wavelengths, 3-dB bandwidths parameters of (450 nm, 40 nm), (550 nm, 40 nm) and (620 nm, 10 nm), respectively. Note that for the experiment, the optical filter spectra are deliberately selected to have an overlap with the deployed LED spectra so the CAOS imaging of the two hole target can respond clearly to the presence or absence of a target spectral response at the given LED illumination spectrum.

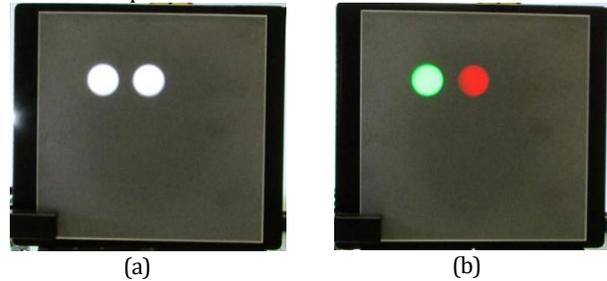

(a)          (b)

Fig.8. Room lighting photo of the deployed test target with 2 holes for testing the active FDMA-CDMA mode. (a) Target with open room white light lit holes and (b) Target with green and red optical filters placed on the 2 different holes.

Figure 8 shows a room lighting illumination standard color camera photo of deployed active FDMA-CDMA mode test target with 2 holes. Fig. 8(a) shows the target with open room white light lit holes and Fig. 8(b) shows the target when green and red optical filters are placed on the 2 different holes. A Thorlabs model DG100X100-220 ground glass 220 GRIT 10 cm×10 cm diffuser sheet (see Fig. 8) is deployed on the target hole surface to create scattered light from the target hole locations. The two holes target is placed at a distance of 56 cm from the L1 imaging lens in the CAOS camera system described earlier. L1 has a 10 cm focal length and is placed 12.2 cm from the DMD. Given that the average CW power from the deployed LEDs is rated over 200 mW creating a bright visible light scenario, the Thorlabs model PDA100A2 variable gain visible band point silicon PD with a gain set at 30 dB is deployed at the PD1 location. To capture the 2 hole target with independent optical filters, the active FDMA-CDMA mode is operated to capture a Q= M×N= 32×15 = 480 CAOS pixels grid. P=3 FDMA channels with a $f_1$, $f_2$, $f_3$ set, J=Q=480 as the 3 LED beams overlap and a W=512 bits CDMA Walsh code is applied with a bit rate of 31 Hz. The CAOS pixel size is 20×20 micromirrors and an ADC sampling rate of 2 Msps with a 10 V max setting for capture of the PD1 photo-detected signal.

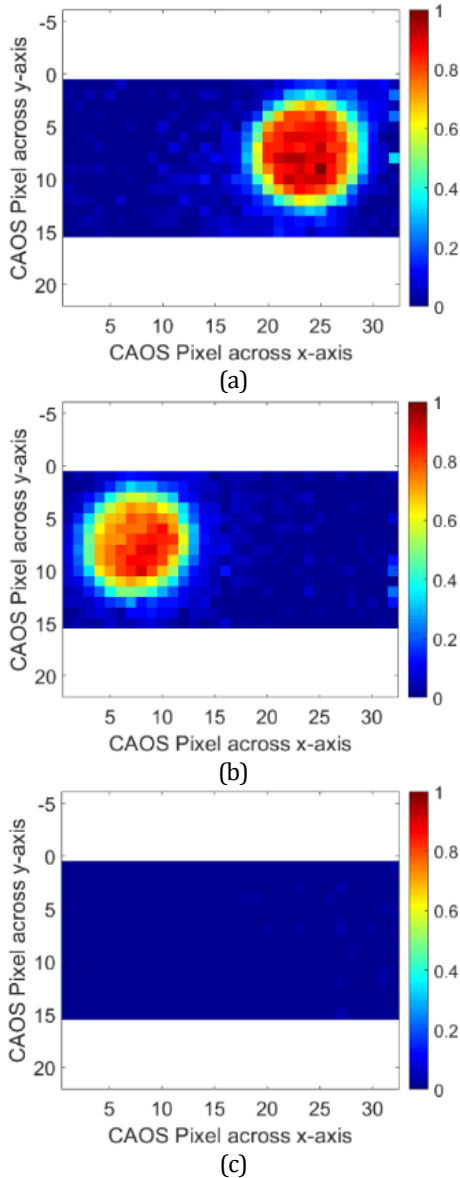

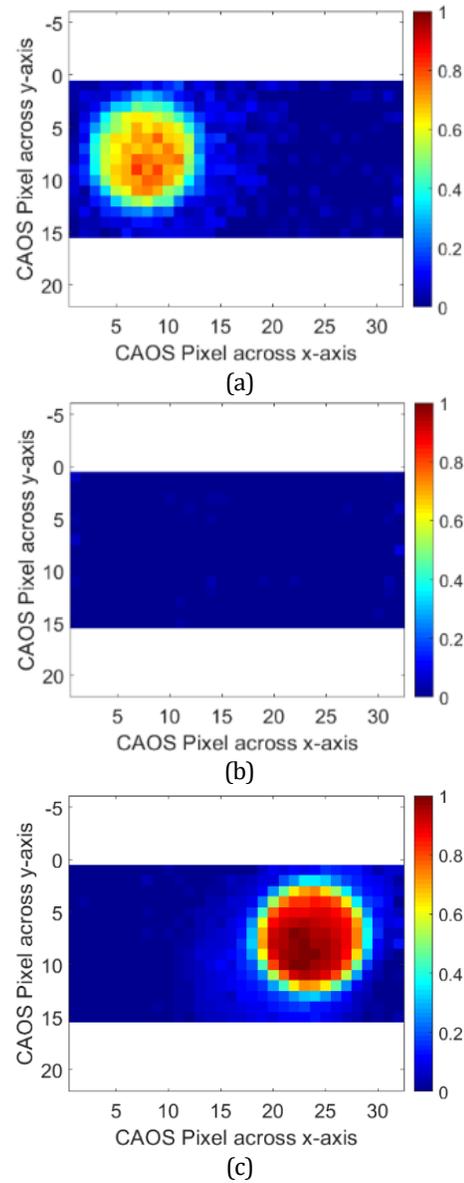

Fig.9. Target designed such that R-filter and G-filter imaged locations are falling on DMD plane left and right sides, respectively. The independent and simultaneous FDMA-CDMA images are shown for (a) LED1 G-light at FM $f_1$=25 KHz, (b) LED2 R-light at FM $f_2$=29 KHz, and (c) LED3 B-light at FM $f_3$=35 KHz.

The first active FDMA-CDMA mode experiment is carried out using a green and red filter on the separate holes of the target with the image of the R filter position and G filter position falling on the left and right sides of the DMD, respectively. The experimental images produced by the active FDMA-CDMA mode for this target scenario are shown in Fig.9(a) for LED1 G-light at FM $f_1$=25 KHz, in Fig.9(b) for LED2 R-light at FM $f_2$=29 KHz, and in Fig.9(c) for LED3 B-light at FM $f_3$=35 KHz. Indeed, the correct target hole light spots show up in the 3 images highlighting the simultaneous optical spectra target response discrimination capabilities of the active FDMA-CDMA mode when using different light source spectra. For example, in Fig.9(a), only the G-light target hole on the right side shows up as the $f_1$=25 KHz encoded image is for LED1 G-light.

Similarly, in Fig. 9(b), the R-light target hole on the left side shows up as $f_2$=29 KHz encoded image is for the LED2 R-light. Furthermore, in Fig.9(c), neither the R-light target hole or the G-light target hole shows up as the $f_3$=35 KHz encoded image is for the LED3 B-light and there is no B-filter placed in any of the target holes. Note that the pixel irradiances shown are normalized by the brightest pixel in the CAOS captured 3-image set to show relative brightness levels of the target spectral responses over the 3 LED spectra.

Fig.10. Target designed such that G-filter and B-filter imaged locations falling on DMD plane left and right sides, respectively. The independent and simultaneous FDMA-CDMA images are shown for (a) LED1 G-light at FM $f_1$=25 KHz, (b) LED2 R-light at FM $f_2$=29 KHz, and (c) LED3 B-light at FM $f_3$=35 KHz.

Figure 10 shows an alternative active FDMA-CDMA mode experiment when the G filter position and B filter position falls on the left and right sides of the DMD, respectively. Again, the correct images of the target holes show up in the 3 simultaneous CAOS images highlighting the target spectra simultaneous multiple light sources response capability of the active FDMA-CDMA mode using overlapping light beams. It is pertinent to point out that the active FDMA-CDMA mode user can tailor the settings of the CAOS camera to meet the user application requirements. For example, the FDMA channel count, CDMA bit rate and FDMA rates can be increased to 10, 22 KHz and 1 MHz, respectively. Such changes can produce much faster frame rates and target spectral images for many more optical spectra. Also note that the 3 LED illumination spot beams through the placed different optical filters have a bit different non-flat non-uniform irradiance levels and this variation can be observed as the different shades of gray-levels observed inside the imaged spot regions shown in Figures 9 and 10.

## 4. CONCLUSION

For the first time designed and demonstrated is the hybrid FDMA-CDMA mode of the CAOS camera. The triple coding FDMA-CDMA mode combines the high SNR feature of the CDMA-mode with the linear HDR and DSP crosstalk filtering capability of the FDMA-mode to form a high security imager with a faster imaging speed than the crosstalk limited FM-CDMA mode. A calibrated 64 dB 6-patches white light HDR target is fully and accurately imaged using the passive FDMA-CDMA mode demonstrating superior image recovery that the FM-CDMA mode. Also demonstrated for the first time is simultaneous imaging of two different spectral bands of a target using the FDMA-CDMA mode. Specifically, a fiber output beam is successfully imaged by the passive FDMA-CDMA mode providing an image over a 350-1000 nm band and another image over an 800 – 1800 nm band. In addition, the active FDMA-CDMA mode is presented that has the capability to simultaneously provide multiple optical spectra tailored target images without the use of a time multiplexed operation tunable filter. This unique core capability of the active FDMA-CDMA mode CAOS camera is experimentally demonstrated using visible light 3 LEDs provided overlapped light beams with unique optical spectra and a two holes spectrally sensitive target designed with 3 different optical filters. The proposed FDMA-CDMA modes of the CAOS camera can be useful for robust linear readings spectrally sensitive NIR light food inspection [33-34] as well as high security bright light imaging applications including fire-fighting and search and rescue.

**Disclosures**. The authors declare no conflicts of interest.

**Data Availability**. Data underlying the results presented in this paper are not publicly available at this time but may be obtained from the authors upon reasonable request.

## References

1. J. E. Shields, R. W. Johnson, M. E. Karr, A. R. Burden, and J. G. Baker, "Daylight visible/NIR whole-sky imagers for cloud and radiance monitoring in support of UV research programs," Proc. SPIE, 5156, 155-166 (2003).
2. G. Verhoeven, "Imaging the invisible using modified digital still cameras for straightforward and low-cost archaeological near-infrared photography," J. Archaeol. Sci. 35 (12), 3087-3100 (2008).
3. C. D. Tran, Y. Cui, and S. Smirnov, "Simultaneous multispectral imaging in the visible and near-infrared region: applications in document authentication and determination of chemical inhomogeneity of copolymers," Anal. Chem. **70**(22), 4701-4708, (1998).
4. B. T. W. Putra and P. Soni, "Evaluating NIR-Red and NIR-Red edge external filters with digital cameras for assessing vegetation indices under different illumination," J. Infrared Phys. & Technol. **81**,148-156, (2017).
5. Y. Kanzawa, Y. Kimura and T. Naito, "Human skin detection by visible and near-infrared imaging," in *IAPR Conference on Machine Vision Applications*, **12** , 503-507 (2011).
6. J. B. Thomas, P.J. Lapray, P. Gouton, and C. Clerc, "Spectral characterization of a prototype SFA camera for joint visible and NIR acquisition," MDPI Sensors, **16**(7:993), (2016).
7. Q. Zhou, Z. Chen, J. Robin, X.-L. Deán-Ben, D. Razansky, "Diffuse optical localization imaging (DOLI) enables noninvasive deep brain microangiography in NIR-II window," Optica, **8**(6), 796-803, (2021).
8. Zephir 1.7 SWIR Infrared Camera Data Sheet, 2019, Photon etc., Montreal, Canada. www.photonetc.com
9. Goldeneye CL-008 Cool TEC1 SWIR Infrared Camera Data Sheet, 2019, Allied Vision, Germany.http://www.alliedvision.com
10. C12741-03 InGaAs camera, 2019, Hamamatsu Photonics, Japan. www.hamamatsu.com
11. OWL 640 SWIR InGaAs camera, 2019, Raptor Photonics, Northern Ireland, UK. www.raptorphotonics.com
12. M. J. E. Golay, "Multi-slit spectrometry," J. Opt. Soc. Am., **39** (6), 437-444, (1949).
13. R. N. Ibbett, D. Aspinall, and J. F. Grainger, "Real-Time Multiplexing of Dispersed Spectra in Any Wavelength Region," Appl. Opt., **7**(6), 1089-1093 (1968).
14. J. A. Decker, Jr. and M. O. Harwitt, "Sequential Encoding with Multislit Spectrometers," Appl. Opt., **7** (11), 2205-2209, (1968).
15. R. A. DeVerse, R. M. Hammaker, W. G. Fateley, "Realization of the Hadamard Multiplex Advantage Using a Programmable Optical Mask in a Dispersive Flat-Field Near-Infrared Spectrometer," J. Appl. Spectroscopy, **54** (12), 1751-1758, (2000).
16. N. A. Riza and S. Sumriddetchkajorn, "Digitally controlled fault tolerant multiwavelength programmable fiber-optic attenuator using a two dimensional digital micromirror device," Opt. Lett., **24** (5), 282-284, (1999).
17. K. Kearney and Z. Ninkov, "Characterization of a digital micro-mirror device for use as an optical mask in imaging and spectroscopy," Proc. SPIE, 3292, 81-92, (1998).
18. S. Sumriddetchkajorn and N. A. Riza, "Micro-electro-mechanical system-based digitally controlled optical beam profiler," Appl. Opt., **41** (18), 3506-3510, (2002).
19. N.A. Riza, S.A. Reza, P.J. Marraccini, "Digital micro-mirror device-based broadband optical image sensor for robust imaging applications," Opt. Commun., **284** (1), 103–111, (2011).
20. D. Takhar, J. N. Laska, M. B. Wakin, M. F. Duarte, D. Baron, S. Sarvotham, K. F. Kelly, and R. G. Baraniuk, "A New Compressive Imaging Camera Architecture using Optical-Domain Compression," Proc. SPIE, 6065, 6065091-10, (2006).
21. M. F. Duarte, M. A. Davenport, D. Takhar, J. N. Laska, T. Sun, K. F. Kelly, & R. G. Baraniuk, "Single-pixel imaging via compressive sampling," IEEE Signal Processing Magazine, **25** (2), 83-91, (2008).
22. N. A. Riza, Coded Access Optical Sensor (CAOS), USA Patent 10356392 B2, 2019.
23. N. A. Riza, M. J. Amin, and, J. P. La Torre, "Coded Access Optical Sensor (CAOS) Imager," J. European Optical Society (JEOS) Rapid Publications, **10**, 150211-8, (2015).
24. N. A. Riza, "Thinking camera—Powered by the CAOS camera platform," Proc. Eur. Opt. Soc. Annu. Meeting (EOSAM), *EPJ Web of Conferences*, 238, 06012 (2020).
25. N. A. Riza and M. A. Mazhar, "177 dB linear dynamic range pixels of interest DSLR CAOS camera," IEEE Photonics J., **11** (3), 1-10, (2019).

26. N. A. Riza and M. A. Mazhar, "Laser beam imaging via multiple mode operations of the extreme dynamic range CAOS camera," Appl. Opt., **57** (22), E20-E31, (2018).
27. J. A. Decker, "Experimental Realization of the Multiplex Advantage with a Hadamard-Transform Spectrometer," Appl. Opt., **10** (3), 510-514, (1971).
28. M. Harwit and N. J. A. Sloane, *Hadamard Transform Optics*, (Academic Press, 1979).
29. T. Mizuno and T. Iwata, "Hadamard-transform fluorescence-lifetime imaging," Opt. Express, **24** (8), 8202-13, (2016).
30. M. Chi, Y. Wu, F. Qian, P. Hao, W. Zhou, and Y. Liu, "Signal-to-noise ratio enhancement of a Hadamard transform spectrometer using a two-dimensional slit-array," Appl. Opt, **56** (25), 7188-7193, (2017).
31. M. A. Mazhar and N. A. Riza, "CAOS spectral imager design and advanced high dynamic range FDMA–TDMA CAOS mode," Appl. Opt, **60** (9), 2488-98, (2021).
32. N. A. Riza and M. A. Mazhar, "The CAOS camera–unleashing the power of full spectrum extreme linear dynamic ranging imaging," *Proceedings of IEEE British and Irish Conf. on Optics & Photonics* (BICOP), 1-4, (2018).
33. H. Huang, L. Liu and M. O. Ngadi, "Recent developments in hyperspectral imaging for assessment of food quality and safety," MDPI Sensors, 14(4), 7248-7276, (2014).
34. X. Fu and and Y. Ying, "Food safety evaluation based on near infrared spectroscopy and imaging: a review," Crit. Rev. Food. Sci. Nutr., **56**(11), 1913-1924, (2016).